# Cumulative Probability of Blast Fragmentation Effect

*Oleg Mazonka, 2012*

***Abstract*** *This paper presents formulae for calculation of cumulative probability of effect made by blast fragments. Analysis with Mott distribution, discrete fragment enumeration, spatial non-uniformity, numerical issues, and a generalisation for a set of effects are also discussed.*

## 1. Introduction

When an explosive device detonates fragments are created out of solid casing. These fragments will have a distribution of masses and spatial dispersion which can be described using probability distributions. The effect of these fragments on a given target depends on the size and terminal velocity of the fragment when it hits the target. So given the probability distribution of effects and the mass and probability distributions of the fragments the question is what is the cumulative probability of an explosion on a target. This paper presents formulae which give the overall probability of a particular effect given the knowledge of the effect probability of one fragment.

There are number of difficult to recognise issues associated with the above problem, which are discussed in detail below. The solutions to those issues given with formally defined mathematical conditions may not always be satisfactory for a reader interested in the subject area due to difference in focus on the relevant problem and preset assumptions. However the list of the discussed questions and the solution techniques would definitely assist anyone interested in this topic.

## 2. Cumulative Probability

We are interested in probability $P$ that one or more fragments of the explosion causes a particular effect by hitting a target. Let us denote the probability of one fragment making the effect, given it hits, as $P_e$, and the probability of hit by one fragment as $F_r$. $P_e$ depends on the mass of the fragment, its terminal velocity, and other external factors such as the aspect angle of the target. Since the final result $P$ depends on the same external factors they can be left out from now on since they are considered to be fixed.

As a simple case let us assume that the terminal velocity $v$ of the fragment depends on the mass of the fragment $m$ and does not depend on other probabilistic physical factors such as, for example, shape factors: $v = v(m)$. In this case $P_e$ depends only on the mass of the fragment: $P_e = P_e(v, m) = P_e(v(m), m) = P_e(m)$.

The mass of a fragment is a random variable; hence it can be described by a probability distribution $n(m)$. This function represents the probability that the fragment's mass is equal to $m$ (not the probability that there exists a fragment with the mass $m$) and is unit normalised:

$$\int_0^\infty n(m)\, dm = 1 .$$

If all the fragments would cause the effect with the same probability $\tilde{P}_e$ assuming hit, then the cumulative probability of the effect is just the probability of one fragment $P_1 = F_r \tilde{P}_e$ summed binomially over all fragments:

$$P = 1 - (1 - P_1)^{N_T} = 1 - \left(1 - F_r \tilde{P}_e\right)^{N_T} \qquad \textbf{1}$$

where $N_T$ is the total number of fragments.

In general the fragments have different masses and terminal velocities hence Eq. 1 can be used only for the number of fragments with the same mass as:

$$P_i = 1 - \left(1 - P_1^i\right)^{N_i} \qquad \textbf{2}$$

Here $P_1^i$ denotes the probability of generating the effect by one of $N_i$ fragments with a mass taken from vicinity $\Delta m$ of mass $m_i$. The total number of such fragments is:

$$N_i = N_T n(m_i)\, \Delta m \qquad \textbf{3}$$

And the final probability will be expressed as:

$$P = 1 - \prod_{i=1}^{\infty} (1 - P_i) \qquad \textbf{4}$$

The index $i$ in Eq 4 above enumerates masses in the range $(0,\infty)$ and it runs to infinity because we safely assume that the distribution $n(m)$ goes zero after some value, for example, definitely after the mass of casing.

Since the mass of a fragment is known, the probability of the effect of the fragment is the probability of hit $F_r$ (by one fragment) multiplied by the probability of the effect $P_e$ given hit:

$$P_1^i = F_r P_e(m_i) \qquad \textbf{5}$$

By inserting Eq 5 and Eq 3 into Eq 2 and then Eq 2 into Eq 4, the cumulative probability becomes:

$$P = 1 - \prod_{i=1}^{\infty} \left(1 - F_r P_e(m_i)\right)^{N_T n(m_i)\Delta m} = 1 - \exp \ln \prod_{i=1}^{\infty} \left(1 - F_r P_e(m_i)\right)^{N_T n(m_i)\Delta m} =$$

$$= 1 - \exp \sum_{i=1}^{\infty} N_T n(m_i)\Delta m \ln\left(1 - F_r P_e(m_i)\right)$$

$$= 1 - \exp N_T \sum_{i=1}^{\infty} n(m_i) \ln\left(1 - F_r P_e(m_i)\right)\Delta m$$

Taking the limit $\Delta m \to 0$ one finds the final formula for $P$:

$$\boxed{P = 1 - \exp N_T \int_0^{\infty} n(m) \ln\left(1 - F_r P_e(m)\right) dm} \qquad \textbf{6}$$

In the case when $F_r P(m)_e << 1$ this equation naturally reduces to the product of average probability of the effect, probability of hit, and the total number of fragments:

$$P = 1 - e^{N_T \int_0^{\infty} n(m) \ln(1 - F_r P_e(m)) dm} = 1 - e^{-N_T F_r \int_0^{\infty} n(m) P_e(m) dm} = N_T F_r \int_0^{\infty} n(m) P_e(m) dm$$

When fragments do not have the effect below some critical mass $m_c$ and have unit effect with masses above the critical mass, i.e. $P_e$ is a Heaviside step function

$$P_e(m) = \mathrm{H}(m - m_c)$$

In this case the integral in Eq 6 can be taken explicitly

$$P = 1 - \exp N_T \int_0^{\infty} dm\, n(m) \ln\left(1 - F_r \mathrm{H}(m - m_c)\right) =$$
$$= 1 - \exp N_T \int_{m_c}^{\infty} dm\, n(m) \ln\left(1 - F_r\right) = 1 - \left(1 - F_r\right)^{N_T \int_{m_c}^{\infty} dm\, n(m)} = 1 - \left(1 - F_r\right)^{N(m_c)} \qquad \textbf{7}$$

where $N(m)$ is a number of fragments with masses greater than $m$:

$$N(m) = N_T \int_m^{\infty} n(m')\, dm' \qquad \textbf{8}$$

Eq. 7 gives the correct expected result for independent $N(m_c)$ fragments of masses $m > m_c$ each hitting the target with the probability $F_r$.

## 3. Mott Equation

The distribution $n(m)$ in Eq 6 is input data for calculation of cumulative probabilities. There are many experimental and theoretical studies dedicated to selection of fragment distribution. One classical and the most popular example is the Mott distribution.

In Ref [1] Sir Nevill F. Mott defines the distribution of explosion fragments as

$$dN = Be^{-\frac{\sqrt{m}}{M_A}} d\sqrt{m}$$

Constant $M_A$ is defined in Ref [1] and depends on physical characteristics of the explosion, such as mass and type of the charge, and mass, type and shape of the casing. The parameter $B$ can be deduced from the total mass of the fragments being equal to the mass of the case $M_0$:

$$dM = m\,dN = mBe^{-\frac{\sqrt{m}}{M_A}} d\sqrt{m}$$

$$M_0 = \int_0^\infty dM = \int_0^\infty Bme^{-\frac{\sqrt{m}}{M_A}} d\sqrt{m} = 2BM_A^3 \quad \text{using the identity} \quad \int_0^\infty x^2 e^{-\frac{x}{y}} dx = 2y^3$$

$$\Rightarrow \quad B = \frac{M_0}{2M_A^3}$$

The number of fragments $N(m)$ shown in Eq 8 can be directly calculated from the above distribution:

$$N(m) = \int_m^\infty dN = \frac{M_0}{2M_A^3} \int_m^\infty e^{-\frac{\sqrt{m'}}{M_A}} d\sqrt{m'} = \frac{M_0}{2M_A^2} e^{-\frac{\sqrt{m}}{M_A}} \qquad \mathbf{9}$$

This result is the well known Mott cumulative distribution of mass fragments. The coefficient in front of the exponent is the total number of fragments:

$$N_T = N(0) = \frac{M_0}{2M_A^2}$$

Combining both Eq 8 and Eq 9 fragment distribution $n$ can be derived:

$$N(m) = N_T \int_m^\infty n(m')\,dm' = N_T e^{-\frac{\sqrt{m}}{M_A}} \quad \Rightarrow$$

$$\int_0^m n(m')\,dm' = 1 - e^{-\frac{\sqrt{m}}{M_A}} \quad \Rightarrow$$

$$n(m) = \frac{\partial}{\partial m}\left(1 - e^{-\frac{\sqrt{m}}{M_A}}\right) = \frac{1}{2M_A\sqrt{m}} e^{-\frac{\sqrt{m}}{M_A}}$$

which is automatically normalised to 1: $\int_0^{\infty} n(m)dm = 1$

The cumulative probability, Eq. 6 can now be written with explicit Mott distribution

$$P = 1 - \exp \frac{M_0}{4M_A^3} \int_0^{\infty} \frac{\ln(1 - F_r P_e(m))}{\sqrt{m}} \exp\left( -\frac{\sqrt{m}}{M_A} \right) dm \qquad \mathbf{10}$$

## 4. Fragment Enumeration

It is interesting to see how the probability calculated with fragment distribution function *n* can also be expressed through a sum over the fragments.

The cumulative Mott formula, Eq. 9, gives the number of fragments above a given mass *m*:

$$N(m) = N_T \exp\left( -\sqrt{m}/M_A \right)$$

Let us enumerate fragments from the heaviest to the lightest assuming that *i*-th fragment corresponding to mass $m_i$. Select $m_i$ so that $m_1$ gives $N(m_1) = 1$, $m_2$ gives $N(m_2) = 2$, and so on:

$$i = N(m_i) = N_T \exp\left( -\sqrt{m_i}/M_A \right)$$

This relation can be rearranged into:

$$m_i = M_A^2 (\ln N_T - \ln i)^2$$

The probability of a particular effect *P* accounted separately for each fragment is

$$P = 1 - \prod_{i=1}^{N_T} (1 - F_r P_e(m_i))$$

or with a less computational error

$$\begin{aligned} P &= 1 - \exp \sum_{i=1}^{N_T} \ln(1 - F_r P_e(m_i)) = \\ &= 1 - \exp \sum_{i=1}^{N_T} \ln\left(1 - F_r P_e\left(M_A^2 (\ln N_T - \ln i)^2\right)\right) \end{aligned} \qquad \mathbf{11}$$

The sum in the above formula can be cut by a maximal index *i* for which $P_e > 0$.

The probability calculated in Eq 11 is systematically underestimated in comparison to the integral form, Eq 10. The assumption was that the fragment has the lower bound mass on the mass scale. A better selection of the mass value would be taking an average over the section on the mass scale:

$$\overline{m}_i = \int_{m_i}^{m_{i-1}} m\,\overline{n}(m)\,dm$$

with $m_0 = \infty$ and the density function $\overline{n}(m)$ identical to *n* but renormalised to 1 on the segment [$m_i$,$m_{i-1}$]. Fortunately this integral can be taken analytically:

$$\overline{m}_i = \int_{m_i}^{m_{i-1}} m\,\overline{n}(m)\,dm = \frac{\int_{m_i}^{m_{i-1}} m\,n(m)\,dm}{\int_{m_i}^{m_{i-1}} n(m)\,dm} = N_T \int_{m_i}^{m_{i-1}} m\,n(m)\,dm =$$

$$= \frac{N_T}{2M_A} \int_{m_i}^{m_{i-1}} m \frac{1}{\sqrt{m}} \exp\left(-\frac{\sqrt{m}}{M_A}\right) dm = \frac{N_T}{2M_A} \int_{\sqrt{m_i}}^{\sqrt{m_{i-1}}} 2x^2 \exp\left(-\frac{x}{M_A}\right) dx =$$

$$= \frac{N_T}{2M_A} 2M_A e^{-\frac{x}{M_A}} \left(x^2 + 2xM_A + 2M_A^2\right)\Big|_{x=\sqrt{m_{i-1}}}^{\sqrt{m_i}} =$$

$$= N_T e^{-\frac{\sqrt{m_k}}{M_A}} \left(m_k + 2\sqrt{m_k} M_A + 2M_A^2\right)\Big|_{k=i-1}^{i}$$

Knowing that

$$\sqrt{m_i} = M_A \ln \frac{N_T}{i}$$

the derivation comes to

$$\overline{m}_i = N_T \frac{k}{N_T} \left( M_A^2 \left( \ln \frac{N_T}{k} \right) + 2M_A^2 \ln \frac{N_T}{k} + 2M_A^2 \right)\Big|_{k=i-1}^{i} =$$

$$= M_A^2 k \left( 1 + \left( 1 + \ln \frac{N_T}{k} \right)^2 \right)\Big|_{k=i-1}^{i} =$$

$$= M_A^2 \left[ i\left( 1 + \left( 1 + \ln \frac{N_T}{i} \right)^2 \right) - (i-1)\left( 1 + \left( 1 + \ln \frac{N_T}{i-1} \right)^2 \right) \right]$$

or

$$\begin{aligned} \overline{m}_i &= M_A^2 (x_i - x_{i-1}) \\ x_i &= i\left(1 + (1 + \ln N_T - \ln i)^2\right) \end{aligned} \qquad \textbf{12}$$

with $x_0 = 0$ because as assumed above $m_0 = \infty$.

And the final enumeration formula is:

$$P = 1 - \exp \sum_{i=1}^{N_T} \ln\left(1 - F_r P_e \left(M_A^2 (x_i - x_{i-1})\right)\right) \qquad \textbf{13}$$

in which $x_i$ is defined by Eq. 12.

The table below shows the numerical comparison of masses of the first ten fragments for the lower-bound (Eq 11) and the average (Eq 13) cases:

| | $N_T$=4000, $M_A$=0.06 | |
|---|---|---|
| $i$ | $m$ | $\overline{m}$ |
| 1 | 0.247649 | 0.314566 |
| 2 | 0.207985 | 0.225258 |
| 3 | 0.186388 | 0.196360 |
| 4 | 0.171781 | 0.178685 |
| 5 | 0.160863 | 0.166089 |
| 6 | 0.152207 | 0.156383 |
| 7 | 0.145076 | 0.148536 |
| 8 | 0.139037 | 0.141979 |
| 9 | 0.133817 | 0.136367 |
| 10 | 0.129232 | 0.131477 |

One can see that, as expected, $m_i < \overline{m}_i < m_{i-1}$, the average mass is greater than the corresponding lower-bound but less than the previous lower-bound mass, and both sequences converge.

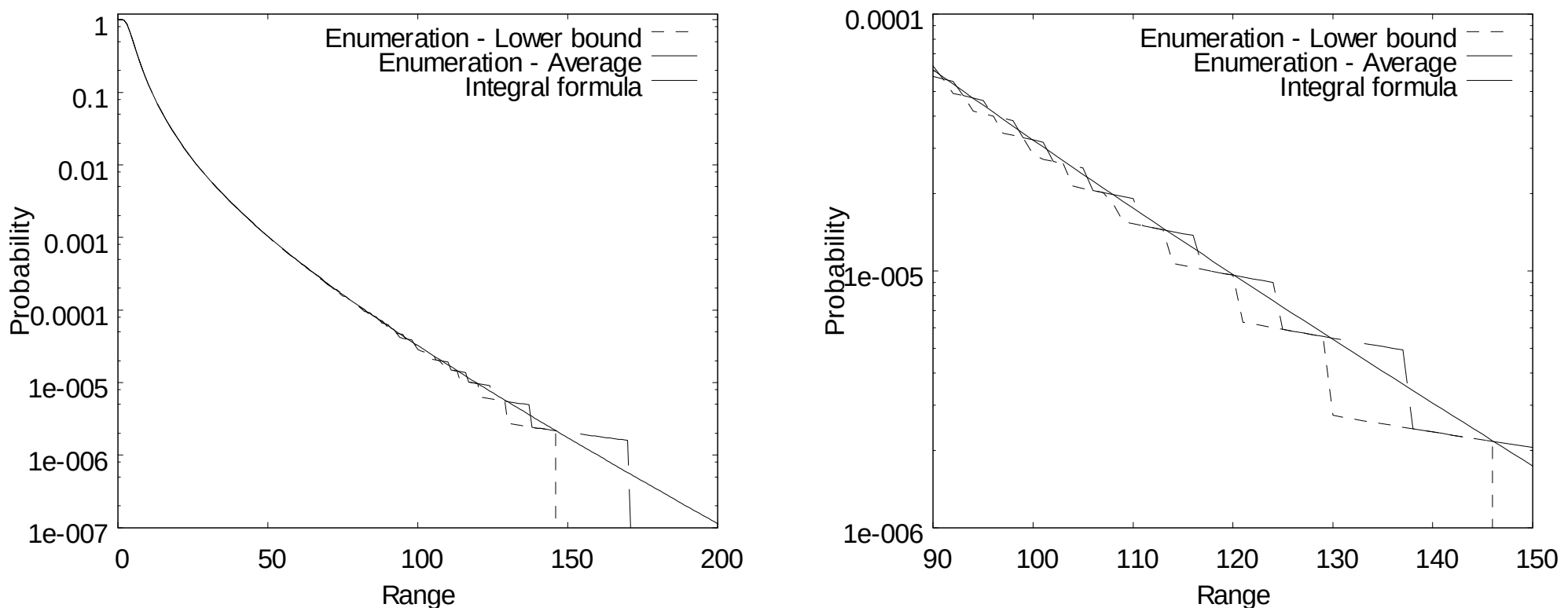

**Figure 1** Probabilities calculated with Eqs **10**, **11**, and **13**.

Figure 1 shows the comparison of calculations with the integral Eq 10**,** lower bound enumeration Eq 11**,** and average enumeration Eq 13. The target is a human size object with a hit effect of fragment's kinetic energy exceeding 80J. The terminal velocity is calculated as $700\exp\left(-0.005r/\sqrt[3]{m}\right)$, where $r$ is a range in meters and $m$ is in $kg$. Parameters of the explosion are: $M_A = 0.024kg^{1/2}$ and $N_T = 1000$. It is interesting to note that enumeration curves look like steps. This is because each new fragment adds its own probability. Slight slope between steps is due to the solid angle decrease as the target viewed from the point of detonation. Lower bound curve touches the integral curve from below, and the average curve goes around the integral as one would predict from the original derivation.

## 5. Geometry Considerations

In the previous sections three results have been obtained: a general formula for cumulative probability, the same formula including Mott mass distributions, and two fragment enumeration formulae. All those results require the value $F_r$, which is the geometric probability a fragment hitting the target.

In the simplest case when all fragments are assured to be thrown uniformly in all directions this value is proportional to the solid angle subtended by the target or the target area under consideration[1]. However this is invalid in the general case of irregular fragment grouping even if the probability distribution of the direction is uniform. This becomes obvious in the extreme case when all fragments are thrown in one (random and uniformly spherically distributed) direction. In this case the probability to hit any convex target cannot be greater than 0.5 (0.5 is half of the sphere when the target is at distance 0 from the point of detonation) whereas uniform distribution of fragments may give the probability close to 1 at short distances depending on $P_e$, and number and distribution of fragments.

To take into account this type of effect let's introduce a parameter $\alpha$ defining the spherical coverage of flying fragments. Let us also assume that geometry of directions forms a solid angle $\Omega_P$ within which the fragment distribution is uniform[2], so that the parameter $\alpha$ changes from 0 – highly asymmetric, all fragments fly in one direction, to 1 – totally symmetric, all fragments fly in all random directions, and

$$\Omega_P = 4\pi\alpha$$

In this case the geometrical probability of hit $F_r$ is

$$F_r(\omega) = \frac{S(\omega)}{\Omega_P}$$

where $S$ is the solid angle of intersection of $\Omega_P$ and the solid angle $\Omega_T$ at which the target is seen from the detonation point. The argument $\omega$ represents the direction of the explosion in relation to the direction to the target from the detonation point. Now $F_r$ depends on $\omega$ – a spherical angle between $\Omega_P$ and $\Omega_T$. This in turn makes $P$ in Eq 6 and its derivatives dependent on this angle: $P = P(\omega)$. If the angle $\omega$ is known when calculating $P$, the final result remains valid for this particular angle. If the angle $\omega$ is unknown, then its distribution has to be defined. In the absence of any additional information the distribution can be assumed to be uniform. So the final result must be recalculated as

$$P_\alpha = \frac{1}{4\pi}\iint P(\omega)\, d\omega \qquad \textbf{14}$$

[1] In the calculations the target can be replaced with some target area surrounding and enclosing the target because the probability to hit the area is always multiplied by the probability of the effect. So the product of those probabilities remains the same.

[2] The symbol Ω is used here and further as both solid angle direction and its absolute value. The appropriate meaning is clear from the context.

In the above formula $\omega$ is a spherical angle hence the integral is double. $P_\alpha$ denotes the cumulative probability differing from $P$ only by taking into account the parameter $\alpha$.

Defining spherical coordinates $\omega=(\varphi,\theta)$ with $\theta$ being the angle between the axes of $\Omega_P$ and $\Omega_T$ and assuming both $\Omega_P$ and $\Omega_T$ being symmetric to rotation around their axes, angle $\varphi$ can be factored out:

$$P_\alpha = \frac{1}{4\pi}\iint P(\omega)\,d\omega = \frac{1}{4\pi}\int_0^{2\pi} d\varphi \int_0^{\pi} P(\theta)\sin\theta\,d\theta = \frac{1}{2}\int_0^{\pi} P(\theta)\sin\theta\,d\theta \qquad \mathbf{15}$$

Then $F_r(\theta)$ and $S(\theta)$ become dependent on $\theta$ only.

To analyse possible values of $S$, let us introduce characteristic apex angles $\theta_P$ and $\theta_T$ by:

$$\Omega_P = 2\pi(1-\cos\theta_P)$$
$$\Omega_T = 2\pi(1-\cos\theta_T)$$

with ranges $\theta \in [0,\pi]$, $\theta_T \in [0,\pi/2]$, $\theta_P \in [0,\pi]$ and two other angles as:

$$\theta_1 = \theta - \theta_P$$
$$\theta_2 = \theta + \theta_P$$

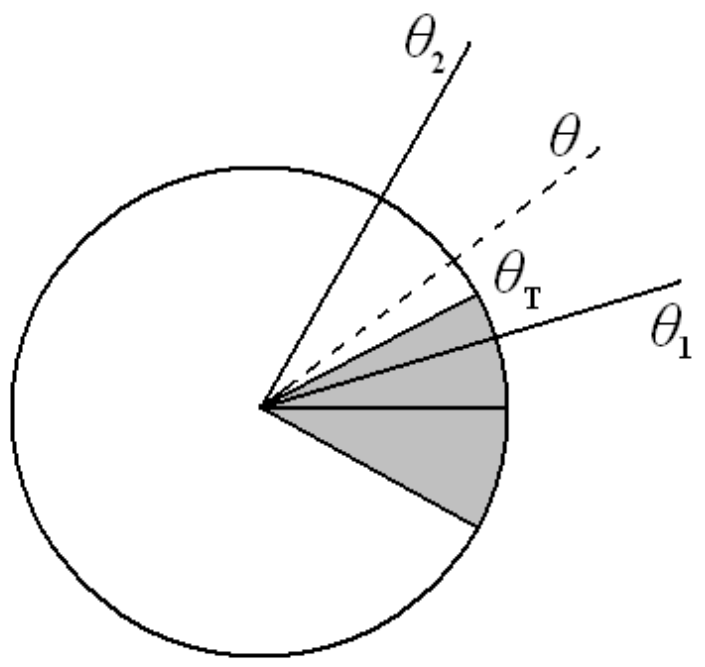

**Figure 2.** Schematic representation of characteristic apex angles of $\Omega_P$ and $\Omega_T$. Shaded area is the target angle, and $\theta_1$ and $\theta_2$ are boundaries of detonation angle.

Figure 2 shows schematically apex angles. All possible configurations of their relative positions are the following:

| Description | Condition | Intersection |
|---|---|---|
| $\Omega_P$ and $\Omega_T$ do not share directions | $\theta_1 > \theta_T$ | $S = 0$ |
| $\Omega_T$ is inside $\Omega_P$ | $\theta_1 < -\theta_T$ | $S = \Omega_T$ |
| Lower bound of $\Omega_P$ is inside $\Omega_T$ and | $-\theta_T < \theta_1 < \theta_T$ | |
| $\Omega_P$ is inside $\Omega_T$ | $\theta_2 < \theta_T$ | $S = \Omega_P$ |
| $\Omega_P$ and $\Omega_T$ intersect | $\theta_T < \theta_2 < 2\pi - \theta_T$ | $S(\theta)$ |
| $(4\pi-\Omega_P)$ is inside $\Omega_T$ | $\theta_2 > 2\pi - \theta_T$ | $S = \Omega_P + \Omega_T - 4\pi$ |

There are no other variants because of the range conditions set above. In the row when $\Omega_P$ and $\Omega_T$ intersect, the solid angle of intersection $S(\theta)$ can take values between zero and $\Omega_P$ or $\Omega_T$ depending on $\theta$ and the difference between $\Omega_P$ and $\Omega_T$.

Analysis of the angle conditions from the table above identifies three distinct cases with each having three distinct ranges for $\theta$:

| Range | **[A]** $\Omega_P < \Omega_T$ | **[B]** $\Omega_P > \Omega_T$ and $4\pi - \Omega_P > \Omega_T$ | **[C]** $\Omega_P > \Omega_T$ and $4\pi - \Omega_P < \Omega_T$ |
|---|---|---|---|
| 1 | $\theta < \theta_T - \theta_P$ | $\theta < \theta_P - \theta_T$ | $\theta < \theta_P - \theta_T$ |
| | $S = \Omega_P$ | $S = \Omega_T$ | $S = \Omega_T$ |
| 2 | $\theta_T - \theta_P < \theta < \theta_T + \theta_P$ | $\theta_P - \theta_T < \theta < \theta_T + \theta_P$ | $\theta_P - \theta_T < \theta < 2\pi - (\theta_T + \theta_P)$ |
| | $0 < S < \Omega_P$ | $0 < S < \Omega_T$ | $\Omega_P + \Omega_T - 4\pi < S < \Omega_T$ |
| 3 | $\theta > \theta_T + \theta_P$ | $\theta > \theta_T + \theta_P$ | $\theta > 2\pi - (\theta_T + \theta_P)$ |
| | $S = 0$ | $S = 0$ | $S = \Omega_P + \Omega_T - 4\pi$ |

Cases [A], [B], and [C] are selected depending on values of $\Omega_P$ and $\Omega_T$. While $\Omega_P$ is fixed and controlled by parameter $\alpha$, $\Omega_T$ depends on the size of the target and distance from detonation point to the target. Values of $S(\theta)$ are represented graphically in Figure 3.

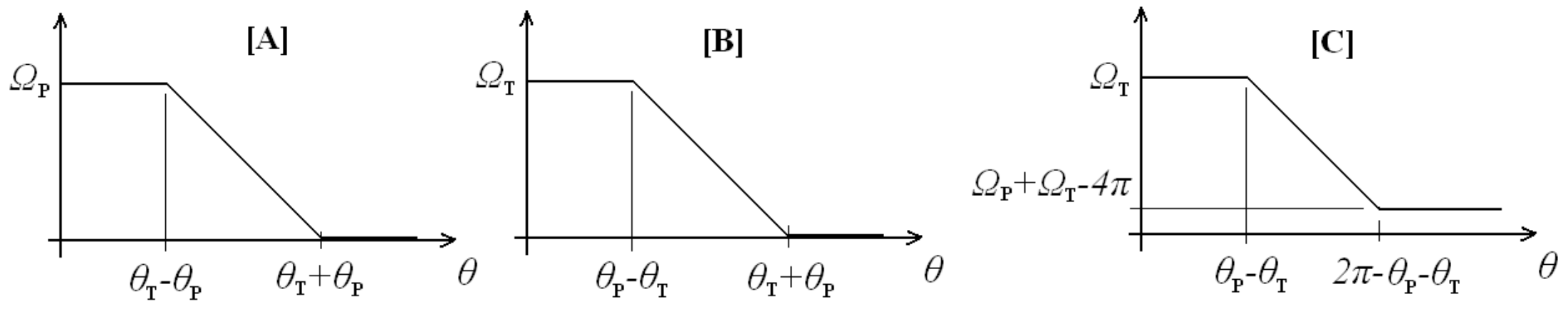

**Figure 3** Schematic representation of $S(\theta)$ for three cases

It is handy to denote the critical values of $\theta$ as:

$$\theta_U = |\theta_T - \theta_P|$$
$$\theta_D = \pi - |\pi - \theta_T - \theta_P|$$

Then using results from the tables above a general solution for $F_r(\theta)$ is:

$$F_r(\theta) = \begin{cases} F_U & \text{if} \quad \theta < \theta_U \\ F_D & \text{if} \quad \theta > \theta_D \\ F_{\alpha\beta}(\theta) & \text{otherwise} \end{cases} \tag{16}$$

with

$$F_U = \begin{cases} 1 & \text{if} \quad \alpha < \beta \\ \beta/\alpha & \text{otherwise} \end{cases}$$

$$F_D = \begin{cases} 0 & \text{if} \quad \alpha + \beta < 1 \\ (\alpha + \beta - 1)/\alpha & \text{otherwise} \end{cases}$$

where $\beta$ is defined by analogy to $\alpha$ as:

$$\Omega_T = 4\pi\beta$$

The function $F_{\alpha\beta}(\theta)$ connects two extreme values $F_U$ and $F_D$. A simple approximation is a linear function:

$$F_{\alpha\beta}(\theta) = F_U + (F_D - F_U)\frac{\theta - \theta_U}{\theta_D - \theta_U}$$

The exact solution however must still be calculated using $S(\theta)$ as:

$$F_{\alpha\beta}(\theta) = \frac{S(\theta)}{4\pi\alpha}$$

The exact solution for $S(\theta)$ is derived in Ref [2]. Numerical calculations (also in Ref [2]) show that a linear function in the middle range gives a very good approximation comparing to the exact solution. On the other hand the exact solution does not introduce extra computational complexity so it can be preferred to its approximated solution.

Having $F_r(\theta)$ calculated the final result can be rewritten including explosion spherical asymmetry parameter using Eq 15. Eq. 6 in this case transforms into:

$$P_\alpha = 1 - \frac{1}{2}\int_0^\pi d\theta \sin\theta \exp N_T \int_0^\infty dm\, n(m) \ln(1 - F_r(\theta) P_e(m)) \qquad \textbf{17}$$

Figure 4 shows a comparison of probabilities calculated with Eq 17 for different values of the parameter $\alpha$. The characteristics of the explosion and the target are the same as in calculations for Figure 1. $\alpha = 0$ corresponds to the case when all fragments fly in the same direction. As expected this curve starts from a value 0.5 at the range 0.

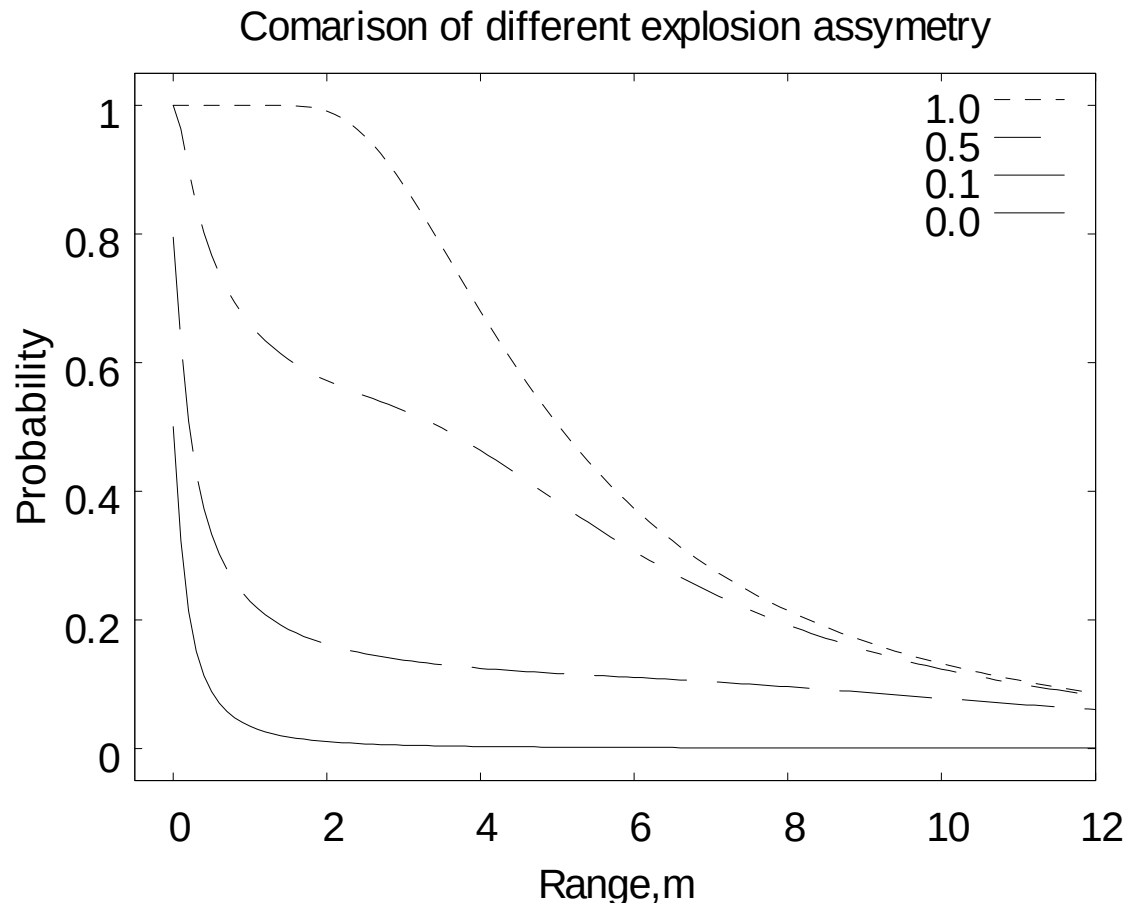

**Figure 4** Probabilities of the effect for a human size target with different values of the parameter $\alpha$.

This analysis gives a good understanding of how blast asymmetry influences the cumulative probability. Here a highly simplistic view has been considered where the symmetry is governed by just one parameter and the fragments are assumed to be uniformly distributed within a cone. The next section generalises the asymmetry to arbitrary spatial distributions.

## 6. Spatial Distribution

Earlier it was assumed that the fragments are distributed according to their masses by the probability distribution $n(m)$ and spatially distributed uniformly in all possible spherical angles, or uniformly within a cone as discussed in Section 5. If information about spatial distribution exists, then Eqs 6 and 17 can be generalised by including distributions in both mass and spherical angles $n(m,\omega)$. This type of generalisation is requires a deeper understanding of the sampling process in the derivation of Eq 6.

First let us consider the original problem with spatially uniform fragment distribution $n(m)$. Below four different sampling algorithms resulting in four different equations presented. Since there are two dimensions: mass $m$ and spherical angle $\omega$, fragments have to be sampled in both $m$ and $\omega$. For each dimension the sampling can be independent for each fragment or assumed that the fragments constitute an ensemble of a particular distribution and each time when a fragment is sampled it is being extracted from the ensemble. Since there are two parameters and two sampling options for each parameter, there are four different combinations. Let us consider each combination separately. As above the following symbols are defined as:

$F_r$ – probability that one fragment hits the target

$P_e$ – probability that a selected fragment causes the effect

$\tilde{P}_e = \int nP_e \, dm$ – probability that one fragment causes the effect

$P$ – cumulative probability

$N_T$ – total number of fragments

$N_i$ – number of fragments with the mass $m_i$

**1) Both mass *m* and spherical angle *ω* are sampled independently**

Let a fragment be selected and its mass is sampled from the probability distribution $n(m)$, and then its spherical angle is sampled from the uniform distribution. The probability that this fragment causes the effect is:

$$P_1 = F_r \tilde{P}_e$$

Now if this process is repeated $N_T$ times, the cumulative probability becomes

$$P = 1 - \left(1 - F_r \tilde{P}_e\right)^{N_T}$$

which is the Eq 1.

This equation does not reduce to a sensible result in the limiting case when $P_e$ is a Heaviside function – zero below a particular critical mass $m_c$ and a unit above:

$$\tilde{P}_e = \int n(m)\mathrm{H}(m - m_c)dn = \frac{N(m_c)}{N_T}$$

$N(m_c)$ is defined by Eq 8. The cumulative probability becomes:

$$P = 1 - \left(1 - F_r \frac{N(m_c)}{N_T}\right)^{N_T}$$

This result is correct from the point of view of the sampling algorithm, but not satisfactory since our final probability must depend only on hit probability $F_r$ and number of fragments above the critical mass $N(m_c)$ and should not depend on total number of fragments, i.e. number of fragments below the critical mass.

**2) Both mass *m* and spherical angle *ω* are selected from an ensemble**

Let all fragments be distributed along the mass scale according to $n(m)$ distribution and uniformly in spatial angle. The probability that a selected with spatial angle $\omega_j$ and mass $m_i$ fragment makes the effect is

$$P_1 = F(\omega_j)P_e(m_i)$$

where $F$ is a function 0 or 1 depending on whether $\omega_j$ is a direction missing or going through the target. Number of fragments of mass $m_i$ is $N_i$, and number of fragments of mass $m_i$ that hit the target is $F_r N_i$, because the fragments are distributed uniformly and $F_r$ is the measure of spherical space occupying by the target. The cumulative probability becomes:

$$P = 1 - \prod_i \prod_j \left(1 - F(\omega_j)P_e(m_i)\right)^{F_r N_i} = 1 - \prod_i \left(1 - P_e(m_i)\right)^{F_r N_i}$$

The product over index $j$ and the function $F$ are cancelled out because the product terms are all reduced to 1 for which $F$=0. The problem with the above result is that deriving the equation analogous to Eq 6 and then taking the limit as in Eq 7, the cumulative probability becomes a binary function depending on whether the expression $F_r N(m_c)$ is greater than 1 or not, that is, again, correct according to the selected sampling method, but not expected as a result.

**3) Mass *m* is sampled and spherical angle *ω* is selected from an ensemble**

In this case the probability of the effect of one fragment hitting the target is exactly

$$P_1 = \tilde{P}_e$$

and the total number of such fragments is $F_r N_T$. The rest of the fragments do not hit the target and do not make the effect, so the cumulative probability is:

$$P = 1 - \left(1 - \tilde{P}_e\right)^{F_r N_T}$$

In the example when $P_e$ is a Heaviside function, this result, as in the first case, reduces to

$$P = 1 - \left(1 - \frac{N(m_c)}{N_T}\right)^{F_r N_T}$$

the equation depending on number of fragments less than critical mass. And the same conclusion as in the first case follows.

**4) Mass *m* is selected from an ensemble and spherical angle *ω* is sampled**

This case is basically presented in Section 2. A fragment of mass $m_i$ is selected from the ensemble $n(m)$ and its spherical angle is sampled from a uniform distribution, so the probability of the effect for one selected fragment is:

$$P_1 = F_r P_e(m_i)$$

Since there are $N_i$ such fragments, the cumulative probability becomes:

$$P = 1 - \prod_i \left(1 - F_r P_e(m_i)\right)^{N_i}$$

which in the limit turns into Eq 6, and in the example of Heaviside function gives Eq 7:

$$P = 1 - \left(1 - F_r\right)^{N(m_c)}$$

that is a mathematical expression for the statement: what is the cumulative probability to hit the target by at least one of $N(m_c)$ independent fragments each having hit probability $F_r$.

The above four examples of different sampling methods show that the "right" sampling algorithm must be chosen as in the fourth case: the mass of a fragment is selected from an ensemble and the spherical angle is sampled from the spherical angle probability distribution. Obviously the calculation is straightforward if there is no correlation between probabilities distribution of mass and spherical angle, i.e. $n(m,\omega) = n(m)n(\omega)$. On the other hand if there are correlations, it is not clear how to pick a fragment with mass from the joint distribution $n(m,\omega)$ – it is easy to sample both $m$ and $\omega$ or to selected from the joint ensemble, but not to select one from an ensemble and to sample the other. Unless we are given the explanation of the physical process causing the correlation between those parameters, there is no (or at least I do not see one) proper way to define the ensemble of mass distribution to be used for calculations.

At this point one can only assume that fragments are selected from the probability distribution of mass ignoring the distribution of spherical angle. In other words all the fragments are collected together in an urn, and then each fragment is picked from the urn and assigned the spherical angle according to the distribution $n(m,\omega)$ given its mass *m*. This distribution according to Bayesian law can be expressed as:

$$n_m(\omega) = \frac{n(m,\omega)}{\int n(m,\omega')d\omega'} \qquad \textbf{18}$$

The integral in the denominator is nothing else but the probability distribution of fragments depending on mass and ignoring spatial distribution:

$$n(m) = \int n(m,\omega)d\omega \qquad \textbf{19}$$

Let *T* be the spherical angular area spanned by the target and $T(\omega_T,\omega)$ is a function which value is 1 if $\omega$ is within *T*, and zero otherwise. The parameter $\omega_T$ is the relative angle between the orientation of the explosion and the direction to the target from the explosion point.

Spatial distribution can also be accompanied by information about distribution of initial velocities of the fragments. If there is a function which uniquely identifies the velocity of the fragment given its mass and spherical angle, then the probability of the effect for one fragment becomes dependent on spherical angle: $P_e = P_e(m,\omega)$. The probability of the effect caused by one fragment is the average value of $P_e$ over all possible spherical angles within the target weighed by spatial distribution:

$$P_1(m,\omega_T) = \int T(\omega_T,\omega)P_e(m,\omega)n_m(\omega)d\omega \qquad \textbf{20}$$

This expression is a consequence that spherical angle is sampled from the distribution $n_m(\omega)$. Using Eqs 2, 3, and 4 and following the same derivation as in Section 2, one obtains the equation for cumulative probability similar to Eq 6:

$$\boxed{P(\omega_T) = 1 - \exp N_T \int_0^\infty dm\, n(m) \ln \int (1 - TP_e) n_m(\omega)d\omega} \qquad \textbf{21}$$

Here arguments for *T* and $P_e$ are the same as in Eq 20 but omitted for convenience. Eq 21 can be rewritten using cross entropy function *H*:

$$\begin{aligned} P(\omega_T) &= 1 - \exp N_T H(S,Q) \\ H(S,Q) &= \int dm\, S \ln Q \\ S &= \int n(m,\omega)d\omega = n(m) \\ Q &= \int (1 - TP_e)n_m(\omega)d\omega \end{aligned} \qquad \textbf{22}$$

It is interesting to see the relation between probability $\tilde{P}_1$ of the effect for one fragment sampled from the distribution and the cumulative probability. Since they are connected by:

$$P = 1 - \left(1 - \tilde{P}_1\right)^{N_T}$$

then

$$\tilde{P}_1 = 1 - \left(1 - P\right)^{1/N_T} = 1 - \exp H\left(S, Q\right)$$

which differs from Eq 22 by only disappearance of $N_T$. The meaning of this equation is that sampling $N_T$ fragments with probability $1 - \exp H(S, Q)$ gives the same result as selecting fragments from the ensemble $n(m)$ and then sampling from $n_m(\omega)$. The same logic applies to Eq 6.

In the limit of uniform spatial distribution and Heaviside function for $P_e$ Eq 21 reduces to Eq 7:

$$\begin{aligned}
P(\omega_T) &= 1 - \exp\, N_T \int_0^\infty dm\, n(m) \ln \int \left(1 - T\,\mathrm{H}(m - m_c)\right) \frac{1}{4\pi} d\omega = \\
&= 1 - \exp\, N_T \int_0^\infty dm\, n(m) \ln\left(1 - \mathrm{H}(m - m_c) \frac{1}{4\pi} \int T\, d\omega\right) = \\
&= 1 - \exp\, N_T \int_{m_c}^\infty dm\, n(m) \ln\left(1 - \frac{\Omega_T}{4\pi}\right) = \\
&= 1 - \exp\, N(m_c) \ln\left(1 - \frac{\Omega_T}{4\pi}\right) = 1 - \left(1 - \frac{\Omega_T}{4\pi}\right)^{N(m_c)}
\end{aligned} \qquad \textbf{23}$$

Two final notes about Eq 21. First, the average over all possible angles $\omega_T$ can be taken in the same way as by Eq 14 when orientation of the explosion relative to the target is unknown. Second, often fragment probability distributions are given axially symmetric. If spherical angle $\omega$ is expressed via spherical coordinates $\theta$ and $\varphi$, and the distribution along $\varphi$ coordinate is uniform, then the probability distribution $n_\theta(m, \theta)$ and $n(m, \omega)$ and their differentials are bound by the relations:

$$\begin{aligned}
n(m, \omega) &= \frac{n_\theta(m, \theta)}{2\pi \sin\theta} \\
d\omega &= \sin\theta\, d\theta\, d\varphi
\end{aligned}$$

because

$$n(m) = \int_0^\pi n_\theta(m, \theta) d\theta = \int_0^{2\pi} d\varphi \int_0^\pi n(m, \omega) \sin\theta\, d\theta = \int n(m, \omega) d\omega$$

Eq 21 can further be simplified for axial symmetric case with an assumption that the spatial distribution function $n_m(\omega)$ and probability of the effect $P_e$ do not change much within solid angle $T$ spanned by the target. Using the conversion

$$n_m(\omega) = \frac{n(m, \omega)}{n(m)} = \frac{n_\theta(m, \theta)}{2\pi \sin\theta\, n(m)}$$

the second integral over $\omega$ can be taken explicitly giving:

$$P(\omega_T)=1-\exp N_T\int_0^\infty dm\, n(m)\ln\left(1-\frac{\Omega_T n_\theta(m,\theta_T)}{2\pi\sin\theta_T\; n(m)}P_e(m,\theta_T)\right) \qquad \textbf{24}$$

In the limit this equation as Eq 23 converges to the same result because for spatial uniform distribution:

$$n_\theta(m,\theta)=\frac{1}{2}n(m)\sin\theta$$

## 7. Complex Targets

Eq 21 in the previous section gives a general result for arbitrary fragment distributions in mass and spatial angle for the fixed position of the target $\omega_T$. Also in the discussions above it has been assumed that the target is a simple object hence the probability of the effect had conventionally been split between the probability of hit and the conditional probability of the effect: terms $F_r P_e$ in Eq 6 and $TP_e$ in Eq 21. If a target represents a complex object, the probability of the effect can be expressed as a function $P_e = P_e(\omega,\omega_T,\phi)$ of spatial angles of the fragment $\omega$, of the target $\omega_T$, and of the target orientation $\phi$, incorporating both probabilities: of hit and of the effect given hit. Since the target is a complex object its position has to be represented by three angles $(\omega_T,\phi)$: two angles for the direction to the target and one angle for its orientation – rotation around the direction line. Eq 21 remains valid in this case with the replacement: $TP_e \to P_e(\omega,\omega_T,\phi)$.

As an example, let us derive the cumulative probability for the conditions outlined in the Section 5: 1) orientation of the explosion is unknown and uniform; and 2) spatial distribution of fragments is uniform within a cone $\alpha$. As in Eq 14 the cumulative probability is taken as an average over the spherical uniform distribution $1/4\pi$ of explosion orientation, which is equivalent to the target direction if viewed from the origin of the explosion. However in this case the target orientation must be included as well with uniform distribution $1/2\pi$ :

$$\begin{aligned}P&=\frac{1}{2\pi}\int_0^{2\pi}d\phi\frac{1}{4\pi}\int d\omega_T P(\omega_T)=\\&=1-\frac{1}{8\pi^2}\int_0^{2\pi}d\phi\int d\omega_T\exp N_T\int_0^\infty dm\, n(m)\ln\int\left(1-P_e(\omega,\omega_T,\phi)\right)n_m(\omega)d\omega=\\&=1-\frac{1}{8\pi^2}\int_0^{2\pi}d\phi\int_0^{2\pi}d\varphi_T\int_0^{\pi}d\theta_T\sin\theta_T\exp N_T\int_0^\infty dm\, n(m)\ln V(\theta,\varphi,\theta_T,\varphi_T,\phi)\end{aligned} \qquad \textbf{25}$$

In the last step in Eq 25 the differentials of the spatial angles were replaced with the differentials of spherical coordinates and the argument in $n_m$ is replaced with a pair of spherical coordinates $n_m(\omega)=n_m(\theta,\varphi)$. Furthermore for simplicity the last double integral over $\omega$ is denoted as a function $V$:

$$V(\theta,\varphi,\theta_T,\varphi_T,\phi)=\int(1-P_e(\omega,\omega_T,\phi))n_m(\omega)d\omega=$$
$$=\int_0^{2\pi}d\varphi\int_0^{\pi}d\theta(1-P_e(\theta,\varphi,\theta_T,\varphi_T,\phi))n_m(\theta,\varphi)\sin\theta$$

Since the spatial fragment distribution is axially symmetric then the probability of the effect depends only on relative angle $\varphi$, i.e. in the above integral the target can be placed at the position $\varphi_T=0$ without changing the value $V$:

$$V(\theta,\varphi,\theta_T,\varphi_T,\phi)=V(\theta,\varphi,\theta_T,\varphi_T=0,\phi)=V(\theta,\varphi,\theta_T,\phi)$$

This means that one integral out of six in Eq 25 can be taken analytically:

$$P=1-\frac{1}{4\pi}\int_0^{2\pi}d\phi\int_0^{\pi}d\theta_T\sin\theta_T\exp N_T\int_0^{\infty}dm\,n(m)\ln V(\theta,\varphi,\theta_T,\phi) \qquad \textbf{26}$$

On the other hand $V$ can be simplified because of the symmetry of $n_m(\theta,\varphi)$. First, it does not depend on $\varphi$: $n_m(\theta,\varphi)=n_m(\theta,)$. Second its value is a constant within a cone with apex angle $\theta_\alpha$ and zero outside:

$$n_m(\theta,)=x\mathrm{H}(\theta-\theta_\alpha)$$

where $x$ is some yet unknown value and H is a Heaviside function. The solid angle of the cone is expressed via its apex angle as: $2\pi(1-\cos\theta_\alpha)$, and by the definition of $\alpha$ it is equal to $4\pi\alpha$, hence:

$$\theta_\alpha=\arccos(1-2\alpha) \qquad \textbf{27}$$

From the normalisation of $n_m(\theta)$ one finds:

$$1=\int_0^{2\pi}d\varphi\int_0^{\pi}d\theta\,n_m(\theta,\varphi)\sin\theta=2\pi x\int_0^{\arccos(1-2\alpha)}d\theta\sin\theta=4\pi x\alpha \quad\Rightarrow\quad x=\frac{1}{4\pi\alpha}$$
$$n_m(\theta,)=\frac{\mathrm{H}(\theta-\theta_\alpha)}{4\pi\alpha}$$

Now the function $V$ can be written as:

$$V(\theta,\varphi,\theta_T,\phi)=\int_0^{2\pi}d\varphi\int_0^{\pi}d\theta(1-P_e(\theta,\varphi,\theta_T,\phi))\frac{\mathrm{H}(\theta-\theta_\alpha)}{4\pi\alpha}\sin\theta=$$
$$=1-\frac{1}{4\pi\alpha}\int_0^{\theta_\alpha}d\theta\,\sin\theta\int_0^{2\pi}d\varphi P_e(\theta,\varphi,\theta_T,\phi)$$

So the final formula reads:

$$P = 1 - \frac{1}{4\pi}\int_0^{2\pi} d\phi \int_0^{\pi} d\theta_T \sin\theta_T \exp\left\{ N_T \int_0^{\infty} dm\, n(m) \ln\left( 1 - \frac{1}{4\pi\alpha} \int_0^{\arccos(1-2\alpha)} d\theta \sin\theta \int_0^{2\pi} d\varphi\, P_e(\theta,\varphi,\theta_T,\phi) \right) \right\} \qquad \mathbf{28}$$

In this equation $P_e$ is obviously depends on $m$ that is not shown explicitly. Eq 28 converges to Eq 17 in case when $P_e$ depends on the angles as $P_e(\theta,\varphi,\theta_T,\phi) = P_e(m)T(\theta,\varphi,\theta_T,\phi)$, where $T$ is a hit-or-miss function as defined in Section 6.

## 8. Numerical Peculiarities

Eq 28 represents the analytical formula for calculation the cumulative probability for the conditions outlined in Section 7. However it does not give straightforward solution when numerical calculations are required. First, the parameter $\alpha$ can legally be zero. How to make sure that the expression under the logarithm is never negative? Second, when the integrals are replaced with numerical sums there is the case when $\theta_T = 0$ and because of sinus function the term makes zero contribution into the sum over $\theta_T$. This is not right, since placing the target in the direction of the explosion should not immediately give unit probability. It is correct analytically though because the weight (or in other words probability) of target having exactly at the angle $(\theta = 0, \varphi = 0)$ out of $4\pi$ is zero.

For simplicity consider fragment enumeration equation analogous to Eq 28:

$$P = 1 - \frac{1}{4\pi}\int_0^{2\pi} d\phi \int_0^{\pi} d\theta_T \sin\theta_T \exp\sum_m^{N_T} \ln\left( 1 - \frac{1}{4\pi\alpha} \int_0^{\arccos(1-2\alpha)} d\theta \sin\theta \int_0^{2\pi} d\varphi\, P_e(\theta,\varphi,\theta_T,\phi) \right) \qquad \mathbf{29}$$

The integral under the logarithm can be replaced with a sum over solid angles spanned by some geometric two-dimensional shapes. For example, it is easy to imagine a rectangular grid covering the target from the detonation point of view with its cells being such shapes. Their solid angles $\omega$ can be calculated by the formula for polyhedral cones in Ref [2]. The whole sphere can be split into two areas: inside the grid: $\sum\omega$, and outside the grid: $4\pi - \sum\omega$. At the same time the sphere can be split into two other areas: inside cone $\alpha$ (denote it with superscript $^{in}$) and outside (denote it with $^{out}$), that is:

$$\sum\omega^{in} + \sum\omega^{out} + \left(4\pi - \sum\omega\right)^{in} + \left(4\pi - \sum\omega\right)^{out} = 4\pi$$

By definition of $\alpha$ :

$$\sum\omega^{in} + \left(4\pi - \sum\omega\right)^{in} = 4\pi\alpha$$

which means that

$$\sum\omega^{in} \le 4\pi\alpha \quad \Rightarrow \quad \sum P_e\omega^{in} \le 4\pi\alpha \qquad \mathbf{30}$$

because $P_e$ is not greater than 1. This inequality means that if the parameter $\alpha$ becomes smaller, the normalisation factor (let's call it $\Omega_p$) cannot become smaller than $\sum \omega^{in}$ because the solid angles $\omega^{in}$ are fixed. In other words when the explosion is narrowly directed it still affects angles $\omega^{in}$ and the result becomes independent of the parameter $\alpha$ once the inequality, Eq 30, is broken. Therefore this normalisation factor $\Omega_p$ can be expressed as:

$$\boxed{\Omega_p = \max\left(4\pi\alpha, \sum \omega^{in}\right)} \qquad \textbf{31}$$

And the double integral expression in Eq 29 can be written in the form:

$$\frac{1}{4\pi\alpha} \int_0^{\arccos(1-2\alpha)} d\theta \, \sin\theta \int_0^{2\pi} d\varphi \, P_e(\theta, \varphi, \theta_T, \phi) \;\; \rightarrow \;\; \Omega_p^{-1} \sum P_e \omega^{in}$$

The integral is replaced with the sum over the solid angles falling inside the cone $\alpha$, that is when angle $\theta$ is less than the apex angle of the cone $\theta_\alpha$, or simply expressed as:

$$\boxed{\omega^{in} = \omega \mathrm{H}(\theta_\alpha - \theta)} \qquad \textbf{32}$$

Now two integrals are left in Eq 29 to be replaced with sums. The first integral is trivial – it can be replaced with a sum over equidistant $N_\phi$ values of $\phi$:

$$\frac{1}{2\pi} \int_0^{2\pi} d\phi \ldots \;\; \rightarrow \;\; \frac{1}{N_\phi} \sum_\phi \ldots$$

The second integral is better to express back to spherical:

$$\frac{1}{2} \int_0^{\pi} d\theta \sin\theta \; f(\theta) = \frac{1}{4\pi} \int_0^{2\pi} d\phi \int_0^{\pi} d\theta \sin\theta \; f(\theta) = \frac{1}{4\pi} \int d\omega \; f(\theta)$$

Here the index $T$ is omitted for simplicity. At this moment the integral can be replaced with a sum over equidistant $N_\theta$ angles $\theta$ or over spherical slices with equal solid angle per slice using Archimedes Hat theorem.

Let $\theta_i = i\pi / N_\theta$ be a set of angles slicing the sphere into $N_\theta$ parts. Each angle $\theta_i$ forms a cone of solid angle $\Omega_i = 2\pi(1 - \cos\theta_i)$. Hence the solid angle of each slice is:

$$\Delta\Omega_i = \Omega_i - \Omega_{i-1} = 2\pi(1 - \cos\theta_i) - 2\pi(1 - \cos\theta_{i-1}) = 2\pi(\cos\theta_{i-1} - \cos\theta_i)$$

Assuming the average angle of the slice to be

$$\theta_T^i = \frac{1}{2}(\theta_i + \theta_{i-1}) \qquad \textbf{33}$$

the integral becomes:

$$\frac{1}{4\pi}\int d\omega\, f(\theta) \quad\rightarrow\quad \frac{1}{4\pi}\sum_i \Delta\Omega_i\, f\left(\theta_T^i\right)=$$
$$=\frac{1}{2}\sum_i\left(\cos\theta_{i-1}-\cos\theta_i\right)f\left(\theta_T^i\right)=\sin\frac{\pi}{2N_\theta}\sum_i \sin\theta_T^i\, f\left(\theta_T^i\right)$$

using trigonometric identity $\cos(a-b)-\cos(a+b)=2\sin a\sin b$.

Getting all found ingredients into Eq 29 the numerical formula becomes:

$$P=1-\frac{1}{N_\phi}\sum_\phi \sin\frac{\pi}{2N_\theta}\sum_\theta \sin\theta_T^i \exp\sum_m^{N_T}\ln\left(1-\Omega_p^{-1}\sum_\omega P_e\omega^{in}\right) \qquad \textbf{34}$$

where $P_e\omega^{in}$ are calculated with $\theta_T^i$ according to Eqs 32 and 33.

Using the other approach – Archimedes Hat theorem – the sphere can be sliced into $N_\theta$ layers of the same thickness $h=2/N_\theta$ and of the same solid angle:

$$\Delta\Omega=2\pi h=\frac{4\pi}{N_\theta}$$

The angle $\theta_i$ of layer *i* satisfies $\cos\theta_i=1-ih$, therefore the average angle $\theta_T^i$ of each layer can be chosen as

$$\cos\theta_T^i=1-\left(i-\frac{1}{2}\right)h=1-\frac{2i-1}{N_\theta}$$

or

$$\theta_T^i=\arccos\left(1-\frac{2i-1}{N_\theta}\right) \qquad \textbf{35}$$

In this case the integral converts into

$$\frac{1}{4\pi}\int d\omega\, f(\theta) \quad\rightarrow\quad \frac{1}{4\pi}\sum_i \Delta\Omega f\left(\theta_T^i\right)=\frac{1}{N_\theta}\sum_i f\left(\theta_T^i\right)$$

and similar to Eq 34

$$\boxed{P=1-\frac{1}{N_\phi}\sum_\phi\frac{1}{N_\theta}\sum_\theta \exp\sum_m^{N_T}\ln\left(1-\Omega_p^{-1}\sum_\omega P_e\omega^{in}\right)} \qquad \textbf{36}$$

with $P_e\omega^{in}$ being calculated by $\theta_T^i$ using Eqs 32 and 35.

Eq 36 seems more accurate with the same number $N_\theta$ because each term in the sum over $\theta_T^i$ is given the same weight. Obviously both Eqs 34 and 36 converge to the same result with larger $N_\theta$.

A great speed up of Eq 36 can be achieved if we notice that the value in the iterations over $\phi$ and $\theta$ is the same for the same pattern of the grid cells $\omega^{in}$. This means that once $\phi$ and $\theta$ are selected, a pattern of $\omega^{in}$ can be obtained, then the exponent expression $\exp\sum_m^{N_T}\ln\left(1-\Omega_p^{-1}\sum_\omega P_e\omega^{in}\right)$ can be calculated and its value stored in a cache table. Next time when other $\phi$ and $\theta$ are selected, but the pattern of $\omega^{in}$ is the same the stored value can be used without recalculating the exponent expression.

Another numerical peculiarity is a computation of the expression

$$P = 1 - \exp\sum\ln(1-x)$$

When $x$ values are small, say less than $10^{-20}$, computer floating arithmetic is unable to calculate $(1-x)$. If all $x$ values are small the expression above numerically gives exactly zero, but analytically:

$$P = 1 - \exp\sum\ln(1-x) \approx 1 - e^{-\sum x} \approx \sum x$$

Since $\ln(1-x) \approx -x - x^2/2$ for small $x$, the expression $\ln(1-x)$ can be replaced with $(-x)$ when $x < 10^{-8}$ assuming numerical precision of "double float" to be around $10^{-16}$. If any of $x$ is close or equal to 1, the sum can be interrupted and the result of the exponent can be assumed to be zero.

## 9. Effect Sets

In all above discussions only one possible effect has been considered and only one value – its cumulative probability – has been calculated. This section extends the discussion to many types of effects.

A naïve and incorrect approach is to use the above derived formulae for different types of mutually exclusive effects and declare the result as the cumulative probabilities of mutually exclusive events for the set of those effects. The problem is that only linear operations can be applied to a vector of probabilities of mutually exclusive events, for example taking means, but not logical operations such as or or and.

The simplest case of two mutually exclusive events, not having or having the effect, can be sought as a vector of two events 0 and 1 correspondingly. Let's denote by $[e]$ the cumulative probability of the event $e$, by $[e]_i$ the probability of the event $e$ caused by the fragment $i$, and by $[e_1e_2]$ the cumulative probability of happening either $e_1$ or $e_2$. In the simplest case of two events {0,1} one finds:

$$[01]=[0]+[1]=1$$
$$[01]_i=[0]_i+[1]_i=1$$

from which follows the cumulative probability for event 1:

$$[0]=\prod[0]_i$$
$$[1]=1-\prod[0]_i=1-\prod(1-[1]_i) \qquad \textbf{37}$$

The above equations mean that the cumulative probabilities and the probabilities for one fragment correspond to events which are mutually exclusive and complete. Eq 37 states that event 1 is *absorbing* in respect to event 0 – in a group of fragments only one fragment with event 1 is enough to cause cumulative event 1. Let's represent this schematically as:

$$0\to 1 \qquad \textbf{38}$$

This rule is the reason of why Eqs 1, 2, and 4 had been constructed in that particular way.

Now let's add another event 2 with a rule

$$0\to 1\to 2$$

meaning that

- if in a group of fragments there is at least one fragment 2, then the cumulative event is 2;
- if in a group of fragments there are no fragments 2 but there is at least one fragment 1, then the cumulative event is 1;
- if the group consists of only fragments 0, then the cumulative event is 0.

This rule implies the following way of calculation cumulative probabilities:

$$[0]=\prod[0]_i$$
$$[01]=\prod[01]_i$$

And since $[01]+[2]=1$ and $[01]=[0]+[1]$,

$$[1]=\prod[01]_i-[0]$$
$$[2]=\prod[012]_i-\prod[01]_i=1-\prod[01]_i$$

These equations give the prescription of calculation cumulative probabilities for events 1 and 2.

Now consider a different situation when event 2 is absorbing of 0 but not absorbing of 1. Each fragment on this case has the set of events {0,1,2}, though the cumulative set of events has to include an *emerging* event 3 which corresponds to having fragments of both 1 and 2 types of events. The rule is schematically represented as:

$$0 \to \begin{Bmatrix} 1 \\ 2 \end{Bmatrix} \to 3$$

So in this case the cumulative probabilities can be calculated in the following way:

$$[0] = \prod [0]_i$$
$$[01] = [0] + [1] \quad \Rightarrow \quad [1] = \prod [01]_i - [0]$$
$$[02] = [0] + [2] \quad \Rightarrow \quad [2] = \prod [02]_i - [0]$$
$$[3] = \prod [012]_i - [0] - [1] - [2] = 1 - [0] - [1] - [2]$$

Adding yet another absorbing event 4 as:

$$0 \to \begin{Bmatrix} 1 \\ 2 \end{Bmatrix} \to 3 \to 4$$

produces

$$[0] = \prod [0]_i$$
$$[1] = \prod [01]_i - [0]$$
$$[2] = \prod [02]_i - [0]$$
$$[3] = \prod [012]_i - [0] - [1] - [2]$$
$$[4] = \prod [0124]_i - \prod [012]_i = 1 - \prod [012]_i = 1 - ([0] + [1] + [2] + [3])$$

And so on for any arbitrary graph of event dependencies.

As an example let's consider a set of the following effects $\{M,F,S,K\}$. This notation is often used for different types of vehicle damage: *M* – mobility kill, *F* – firepower kill, *S* – mobility and firepower kill, and *K* – catastrophic kill. Let *N* be no effect event, *U* be *S* effect produced by *M* and *F* fragments but no *S* fragments, and *V* be *S* effect produced by *S* fragment. Then the event set for fragments is $\{N,M,F,V,K\}$ and the cumulative event set is $\{N,M,F,U,V,S,K\}$ with $[U] + [V] = [S]$. The diagram for these events is:

$$N \to \begin{Bmatrix} M \\ F \end{Bmatrix} \to U \to V \to K \qquad \textbf{39}$$

And as before, the cumulative probabilities can be sequentially calculated as:

$$
\begin{aligned}
&[N]=\prod[N]_i\\
&[M]=\prod[NM]_i-[N]\\
&[F]=\prod[NF]_i-[N]\\
&[U]=\prod[NMF]_i-[N]-[M]-[F]\\
&[V]=\prod[NMFV]_i-\prod[NMF]_i=\prod[NMFV]_i-([N]+[M]+[F]+[U])\\
&[S]=[U]+[V]=\prod[NMFV]_i-([N]+[M]+[F])\\
&[K]=1-\prod[NMFV]_i=1-([N]+[M]+[F]+[S])
\end{aligned}
\qquad \mathbf{40}
$$

Now the question is: if the rule of Eq 38 generates the prescription of Eq 37 which in turn results, for example, in Eq 36, then what result produces the prescription of Eq 40 generated by rule of Eq 39 ?

First note that the expression under logarithm in Eq 36 is $[0]_i$ and the cumulative probability is [1]. So Eq 36 can be rewritten as:

$$
[1]=1-\frac{1}{N_\phi}\sum_\phi\frac{1}{N_\theta}\sum_\theta\exp\sum_i\ln(1-[1]_i)=1-[0] \qquad \mathbf{41}
$$

Comparing Eqs 41 and 37 one deduces the following correspondence:

$$
\begin{aligned}
&[x]_i \quad \to \quad \Omega_p^{-1}\sum_\omega P_e^{[x]}\omega^{in}\\
&\prod[x]_i \quad \to \quad \frac{1}{N_\phi}\sum_\phi\frac{1}{N_\theta}\sum_\theta\exp\sum_i\ln[x]_i
\end{aligned}
\qquad \mathbf{42}
$$

Here $P_e^{[x]}$ is the probability that the fragment causes the event *x*. Note that the event *x* in Eq 42 cannot be an event outside the grid; otherwise the expression must include the angle within $\Omega_p$. Assuming that only the event *N* happens outside the grid and it happens with probability 1, the expression in Eq 42 has to be written as:

$$
[N]_i \quad \to \quad \Omega_p^{-1}\left(\sum_\omega P_e^{[N]}\omega^{in}+1\cdot\left(\Omega_p-\sum_\omega\omega^{in}\right)\right)=1-\Omega_p^{-1}\sum_\omega\left(1-P_e^{[N]}\right)\omega^{in}
$$

The event *x* may be a combination of several events in which case it is an or'ed probability of either event. Since we consider only mutually exclusive events, this probability is the arithmetic sum of probabilities of all events caused by the fragment. The meaning of Eq 42 is that substitution of terms in Eq 37 according to Eq 42 produce Eq 36. By analogy, substitution of terms in Eq 40 will result in equations required for calculations cumulative probabilities of the rule defined by Eq 39:

$$
\begin{aligned}
&P^{[0]}=[N]=\prod[N]_i=\frac{1}{N_\phi}\sum_\phi\frac{1}{N_\theta}\sum_\theta\exp\sum_m^{N_T}\ln\left(1-\Omega_p^{-1}\sum_\omega\left(1-P_e^{[N]}\right)\omega^{in}\right)\\
&P^{[1]}=\prod[NM]_i=\frac{1}{N_\phi}\sum_\phi\frac{1}{N_\theta}\sum_\theta\exp\sum_m^{N_T}\ln\left(1-\Omega_p^{-1}\sum_\omega\left(1-P_e^{[NM]}\right)\omega^{in}\right)\\
&\ldots
\end{aligned}
$$

The effects combined with $N$ event can be expressed by the sum of the rest of the effects since the set is collectively exhaustive:

$$1 - P_e^{[N]} = P_e^{[M]} + P_e^{[F]} + P_e^{[V]} + P_e^{[K]} = P_e^{[MFVK]}$$
$$1 - P_e^{[NM]} = P_e^{[F]} + P_e^{[V]} + P_e^{[K]} = P_e^{[FVK]}$$
$$\ldots$$

The iterations can be done simultaneously at once. So the final result of Eq 40 in obtaining the form of Eq 36 is:

$$\begin{Bmatrix} P^{[0]} \\ P^{[1]} \\ P^{[2]} \\ P^{[3]} \end{Bmatrix} = \frac{1}{N_\phi}\sum_\phi \frac{1}{N_\theta}\sum_\theta \exp\sum_m^{N_T} \ln\left(1 - \Omega_p^{-1}\sum_\omega \begin{Bmatrix} P_e^{[MFVK]} \\ P_e^{[FVK]} \\ P_e^{[MVK]} \\ P_e^{[K]} \end{Bmatrix} \omega^{in}\right)$$
$$P^{[M]} = P^{[1]} - P^{[0]}$$
$$P^{[F]} = P^{[2]} - P^{[0]}$$
$$P^{[S]} = P^{[3]} + P^{[0]} - P^{[1]} - P^{[2]}$$
$$P^{[K]} = 1 - P^{[3]} \quad \textbf{43}$$

## 10. Conclusion

In this paper a generic formula, Eq. 6, and formula with Mott distribution Eq. 10 for cumulative probability have been derived. Fragment enumeration estimate is calculated for two cases and compared with the integral formula. The comparison showed a good agreement and the behaviour of intuitive prediction. Analysis is made for asymmetric explosions. Eqs 16-17 show how calculations can be adjusted with consideration of the extra parameter for spherical coverage for fragments, and Eq 21 is derived for the general case with fragment spatial distribution. Cumulative probability with a complex target and uniform distributions of fragments within a cone, Eq 28, has been derived as an example. The last two sections covered numerical aspects of cumulative probability calculations and generalisation for several types of effects.

The solutions presented in this paper give insight on pitfalls associated with a seemingly simple problem of calculating cumulative probabilities of fragmentation effects. The discussions and the results can assist in solving similar problems.

The solutions in this paper have been incorporated into models supporting vulnerability and lethality data required for combat simulations, Ref [3]. These models are aimed to be simple enough to accommodate calculation of many types of weapon and targets, along with the complex interactions between them. The fidelity of these models is bound to timeframes and availability of physical data, representing a trade-off between the speed of data generation and the accuracy of resulting information. From this point of view the solutions discussed in this paper can be elaborated and enhanced depending on the level of details of formulated problem.